# A layer-stress learning framework universally augments deep neural network tasks

Shihao Shao[1*], Yong Liu[2], Qinghua Cui[1*]

[1]Department of Biomedical Informatics, MOE Key Lab of Cardiovascular Sciences, School of Basic Medical Sciences, Peking University, 38 Xueyuan Rd, Beijing, 100191, China

[2]School of Artificial Intelligence, Beijing University of Posts and Telecommunications, 10 Xitucheng Rd, Beijing, 100876, China

* To whom correspondence should be addressed:

Shihao Shao, Email: shaoshihao@pku.edu.cn

Qinghua Cui, Email: cuiqinghua@hsc.pku.edu.cn

Deep neural networks (DNN) such as Multi-Layer Perception (MLP) and Convolutional Neural Networks (CNN) represent one of the most established deep learning algorithms. Given the tremendous effects of the number of hidden layers on network architecture and performance, it is very important to choose the number of hidden layers but still a serious challenge. More importantly, the current network architectures can only process the information from the last layer of the feature extractor, which greatly limited us to further improve its performance. Here we presented a layer-stress deep learning framework (*x*-NN) which implemented automatic and wise depth decision on shallow or deep feature map in a deep network through firstly designing enough number of layers and then trading off them by Multi-Head Attention Block. The *x*-NN can make use of features from various depth layers through attention allocation and then help to make final decision as well. As a result, *x*-NN showed outstanding prediction ability in the Alzheimer's Disease Classification Technique Challenge PRCV 2021, in which it won the top laurel and outperformed all other AI models. Moreover, the performance of *x*-NN was verified by one more AD neuroimaging dataset and other AI tasks.

**Introduction**

Deep neural networks (DNN) such as Multi-Layer Perception (MLP) and Convolutional Neural Networks (CNN)[1] together with their derived architectures such as EfficientNet [2] and Swin Transformer [3] represent the most fundamental, best-known, most widely-used, and most powerful algorithms among various deep learning models [4,5]. Here we take MLP as an example. Basically, MLP consists of an input layer to receive the input data, an output layer to make a prediction about the input data, and multiple hidden layers in between the above two to perform nonlinear transformation of the input data. The number of hidden layers is one of the two hyperparameters of a MLP that determine the overall network architecture and have tremendous effects on network performance. Empirically, increasing the number of hidden layers may improve the representational quality but bigger numbers of hidden layers can have harmful outcomes on the overall network performance in generalizing to new observations [6]. It was reported that networks with more hidden layers frequently performed worse [7]. Therefore, deciding the number of hidden layers in a MLP is very important and thus should be cautiously considered. A number of approaches have been developed for this issue, however, it is still a very confusing and daunting challenge [8]. In the other hand, a DNN architecture for classification tasks can be normally divided into 2 sub-structures: the feature extractor and the classification head. The current DNN architectures can only process the information from the last layer of the feature extractor but missed the information from other depth layers, which, however, we hypothesized can further improve the classification head.

Here we presented a deep neural network layer-stress learning framework based on Multi-Head Attention Block [9]. The basic idea is that the framework first designs enough number of layers for a specific task. Then, it implemented automatic and wise depth decision on shallow or deep feature map in a DNN (e.g. MLP or other architectures) through progressively trading off the layers by Multi-Head Attention Block. In other words, the proposed framework is a form of classification head in regarding of the feature extractor - classification head structure, which thus can gather and process various level features from different depth layers by the Multi-Head Attention Block. Given that the novel framework is applicable to DNN architectures, here we named it as *x*-NN. As a result, if the task only needs shallow depth feature, then the attention will be highly allocated to those shallow feature maps, or they will be given to deeper feature maps.

To test the performance of the *x*-NN framework, we first applied it to the diagnosis classification of Alzheimer's disease (AD), the most common type of dementia and the $5^{th}$ leading cause of mortality worldwide [10]. The precise diagnosis of AD is thus very important to slow and prevent its progression but still faces big challenges [11]. Recently, deep learning have been applied to the diagnosis of AD using neuroimaging data [12], however, they still have obvious debate and have high false positives and false negatives as well in the clinical practice of AD[11]. Therefore, AI models with ease-of-use and better performance for AD diagnosis are emergently needed but still a difficult journey. As a result, *x*-NN showed outstanding prediction ability in the Alzheimer's Disease Classification Technique Challenge PRCV 2021, in which it won the top laurel and outperformed all other AI models. Moreover, the performance of *x*-NN was verified by one more AD neuroimaging dataset from Kaggle. In addition, the performance of *x*-NN was further validated in the prediction of admission to the ICU of confirmed COVID-19 cases.

**Methods and Materials**

**The deep neural network layer-stress learning framework**

The inspiration of the proposed deep neural network layer-stress learning framework ($x$-NN) is from Multi-Head Attention Block, which was originally designed to identify the correlations among feature vectors in the input data. Firstly, $x$-NN designs enough number of hidden layers for a given DNN architecture and then Multi-Head Attention Block is presented to weight each layer, that is, to give several depth-level feature map attentions for all layers. Therefore, if the task only needs shallow depth feature, then the attention will be highly allocated to those shallow feature maps, or they will be given to deeper feature maps. Moreover, Sigmoid functions are applied in the hidden layer particularly to reduce the noise in feature vectors fed to Multi-Head Attention Block.

**Datasets used in this study**

Here two datasets for AD diagnosis were used. One is from the Alzheimer's Disease Classification Technique Challenge PRCV 2021 (https://www.huaweicloud.com/zhishi/PRCV2021.html), which contains 2600 samples and each sample include 28169 features. This dataset includes 781 samples with AD, 1148 samples with Mild cognitive impairment (MCI) and 671 normal controls (NC). Each sample consists of 28169 MRI-derived features (https://ma-competitions-bj4.obs.cn-north-4.myhuaweicloud.com/ad/npy_data_explain.zip).

The other is from the MRI and Alzheimers Magnetic Resonance Imaging Comparisons of Demented and Nondemented Adults (https://www.kaggle.com/jboysen/mri-and-alzheimers). Its MRI data comes from the Open Access Series of Imaging Studies (OASIS)

[https://www.oasis-brains.org/]. It covers MRI data from the elder patient aged 60 to 96, including 363 samples. We randomly assigned 80% of the samples as training dataset and 20% as the validation dataset. Moreover, we further tested the predictive ability of *x*-NN on the prediction of admission to the ICU of confirmed COVID-19 cases using clinical data.

## RESULTS

**The deep neural network layer-stress learning framework**

Here we introduce the presented deep neural network layer-stress learning framework (*x*-NN) in more details. As shown in Figure 1, the input is in the shape of (L, ). Following the input layer, there are *k* Basic Layers connecting one by one to the downstream. Each Basic Layer generates 2 feature vectors, one will be fed into the next Basic Layer, whereas the other will be fed into the Multi-Head Attention Block. From the second output of the Basic Layers, we can get *k* feature vectors. Firstly, all the feature vectors will go through the Linear Layers to keep their shape the same. Then, Sigmoid functions are applied to reduce their noise. After that, those feature vectors will be stacked by a concatenating layer. Each feature vector stacked before, is fed into the Multi-Head Attention Block to calculate $Q*K^T$, which further multiply the V matrix of this block after the scaling. From the Multi-Head Attention Block, we can get feature map in the same shape of the input of this block. We then flatten the feature map into a single feature vector. It will go through a Linear Layers and get the final output.

The critical part of this model can be described as below:

$$\text{Output} = s(FC(\text{LeakyReLU}(FC(\text{Flatten}(\text{MultiHeadAttention}(Q,K,V))))))$$

where:

$$Q, K, V = Concat(s(FC(V1)), s(FC(V2)),..., s(FC(Vn)))$$

**Diagnostic classification of Alzheimer's disease**

For the task of the Alzheimer's Disease Classification Technique Challenge PRCV 2021, *x*-NN designed a MLP with 3 Basic layers including 9 hidden layers. As a result, from epoch of ~10, the loss of *x*-NN is significantly lower than that of the control without fitting (Figure 2). While from epoch of ~15, the F1-Score is significantly greater than that of the control without overfitting. For the MRI and Alzheimers Magnetic Resonance Imaging Comparisons of Demented and Nondemented Adults, *x*-NN obtained similar results. *x*-NN decreased the loss by 1.04% (loss from 0.1506 to 0.1402) and increase the accuracy by 1.89%. These results suggest that the *x*-NN has a significantly higher accuracy and better performance than the control without fitting.

**Prediction of admission to the ICU of confirmed COVID-19 cases**

Coronavirus disease (COVID-19) is an infectious disease caused by the SARS-CoV-2 virus[13]. AI technique has been applied to several important aspects of COVID-19 including drug repurposing[14], mortality prediction[15], radiographic COVID-19 detection[16] etc, among which the prediction task of the admission to the ICU of confirmed COVID-19 cases is still an important and challenging problem. Using data from Kaggle, *x*-NN showed significantly increased performance, including lower loss (Figure 4) and higher AUC than the control (Figure 5) without fitting, suggesting *x*-NN is more stable, reliable, and accurate than the control.

**The attention distribution among the layers**

The above results confirmed that the presented *x*-NN framework has better performance than the control. As described above, *x*-NN augmented deep neural networks through weighting each layer (i.e. giving several depth-level feature map attentions for all layers using Multi-head Attention Block). It is thus interesting to explore the attention distribution among the layers in a specific task. To do so, here we investigated the attention distribution of the eight-head of the Multi-Head Attention Block in the task of the Alzheimer's Disease Classification Technique Challenge PRCV 2021. As shown in Figure 6, each heatmap consists of a 3*3 heatmap block, where the *x*-axis and the *y*-axis represent the donors and the receptors of attention, respectively. It is clearly that the heatmaps with the same *y*-axis value are closer with other. This suggests that as expected the attention is indeed correlated with the network depth and the *x*-NN can adaptively allocate attention according to layers.

**DISCUSSION**

In a summary, we presented a novel deep neural network learning framework, *x*-NN, which implemented automatic and wise depth decision on shallow or deep feature map in a deep neural architecture (e.g. MLP and CNN) through firstly designing enough number of layers and then Multi-Head Attention Block will automatically trade off them. Moreover, *x*-NN can be considered to be a form of classification head in regarding of the feature extractor - classification head structure, which thus can gather and process various level features from different depth layers by the Multi-Head Attention Block. Next, the performance of *x*-NN was confirmed by three case studies. In addition, we found that *x*-NN can also augment deep learning in image data based task, for example the Flower Classification with TPUs (https://www.kaggle.com/c/flower-classification-with-tpus, data not shown). In this study, we implemented and tested *x*-NN in MLP. It should be emphasized that, as a classification head,

*x*-NN is not limited to MLP but can be put into different feature extractors. Actually, *x*-NN can be easily applied or extended to other deep neural networks, such as CNN and Swin Transformer.

For CNN, however, we should dig further at the way how we can feed the feature maps to the Multi-Head Attention Block, rather than feature vectors. Moreover, if we want to use *x*-NN to process features and apply them to various computer vision tasks (e.g. segmentation, detection), we should find a way to better save the sparse information through the processing. For Swin-Transformer, in contrast, *x*-NN can keep the spatial information easily due to the sliding window described in Swin-Transformer already contains spatial information.

In fact, *x*-NN can be seen as a classification head to replace the traditional approach (e.g. FC followed by Global Average Pooling). The tricky thing is, unlike what the traditional one does, *x*-NN can be a part of the decoder in the segmentation and detection network.

Finally, this study presented a general augmented deep neural network learning framework, which is expected to achieve more successful applications in various AI tasks in the future.

**REFERENCES**

1   LeCun, Y., Bottou,L., Bengio, Y., Haffner, P. Gradient-based learning applied to document recognition. *Proceedings of the IEEE* (1998).
2   Tan, M., Le, Q.V. EfficientNet: Rethinking Model Scaling for Convolutional Neural Networks. *International Conference on Machine Learning* (2019).
3   Liu, Z., Lin, Y., Cao, Y., Hu, H., Wei, Y., Zhang, Z., Lin, S., Guo, B. Swin Transformer: Hierarchical Vision Transformer using Shifted Windows. *arXiv:2103.14030* (2021).
4   Kruse, R., Borgelt, C., Branue, C., Mostaghim, S., Stenbrecher, M. Computational intelligence: a metholological introduction (2nd edition). *Springer-Verlag London* (2016).

5   Yamashita, R., Nishio, M., Do, R. K. G. & Togashi, K. Convolutional neural networks: an overview and application in radiology. *Insights into imaging* **9**, 611-629, doi:10.1007/s13244-018-0639-9 (2018).

6   Bisong, E. Building Machine Learning and Deep Learning Models on Google Cloud Platform: A Comprehensive Guide for Beginners. *Apress* (2019).

7   Bengio, Y., LeCun, Y. Scaling learning algorithms towards AI. *Large-Scale Kernel Machines* **1**, 1-41 (2007).

8   Choldun R, M., Santoso, J., Surendro, K. Determining the Number of Hidden Layers in Neural Network by Using Principal Component Analysis. *Intelligent Systems and Applications*, 490-500 (2019).

9   Vaswani, A., Shazeer, N., Parmar, N., Uszkoreit, J., Jones, L., Gomez, A. N., ⋯ Polosukhin, I. Attention is all you need. *Advances in neural information processing systems*, 5998–6008 (2007).

10  Dumurgier, J. & Sabia, S. [Epidemiology of Alzheimer's disease: latest trends]. *La Revue du praticien* **70**, 149-151 (2020).

11  Dubois, B. *et al.* Clinical diagnosis of Alzheimer's disease: recommendations of the International Working Group. *The Lancet. Neurology* **20**, 484-496, doi:10.1016/S1474-4422(21)00066-1 (2021).

12  Jo, T., Nho, K. & Saykin, A. J. Deep Learning in Alzheimer's Disease: Diagnostic Classification and Prognostic Prediction Using Neuroimaging Data. *Frontiers in aging neuroscience* **11**, 220, doi:10.3389/fnagi.2019.00220 (2019).

13  Initiative, C.-H. G. Mapping the human genetic architecture of COVID-19. *Nature*, doi:10.1038/s41586-021-03767-x (2021).

14  Pham, T. H., Qiu, Y., Zeng, J., Xie, L. & Zhang, P. A deep learning framework for high-throughput mechanism-driven phenotype compound screening and its application to COVID-19 drug repurposing. *Nature machine intelligence* **3**, 247-257, doi:10.1038/s42256-020-00285-9 (2021).

15  Banoei, M. M., Dinparastisaleh, R., Zadeh, A. V. & Mirsaeidi, M. Machine-learning-based COVID-19 mortality prediction model and identification of patients at low and high risk of dying. *Critical care* **25**, 328, doi:10.1186/s13054-021-03749-5 (2021).

16  DeGrave, A. J., Janizek, J. D. & Lee, S. I. AI for radiographic COVID-19 detection selects shortcuts over signal. *medRxiv : the preprint server for health sciences*, doi:10.1101/2020.09.13.20193565 (2020).

**Figure Legends**

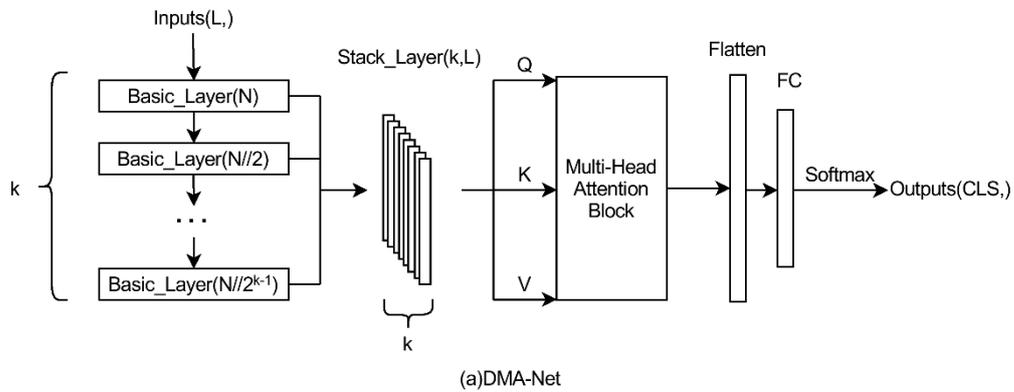

(a)DMA-Net

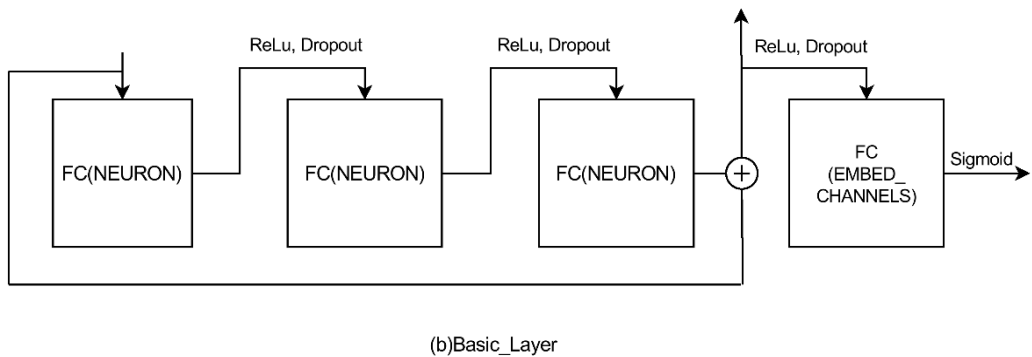

(b)Basic_Layer

Figure 1: The Layer-Stress Learning Architecture — (a) represents the structure of the Layer-Stress Learning Network. N, N//2 , ..., N//(2k-1) denote Neuron in the Basic_Layer. TheStack_Layer(k, L) means stacking the input data to the shape of (k, L). (b) is the structure of the Basic_Layer. NEURON is the neuron number of a certain layer.

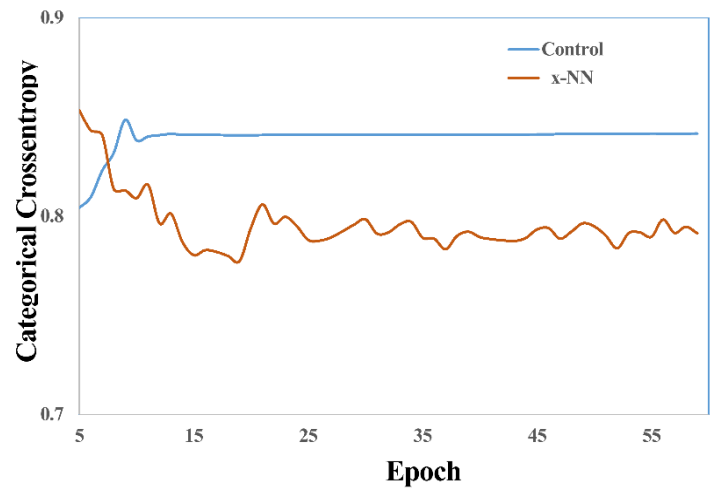

Figure 2. The categorical crossentropy loss of the *x*-NN network with Epoch on the dataset from Alzheimer's Disease Classification Technique Challenge PRCV 2021. It obviously that from epoch of ~10, the *x*-NN has a significantly lower loss than the control without fitting.

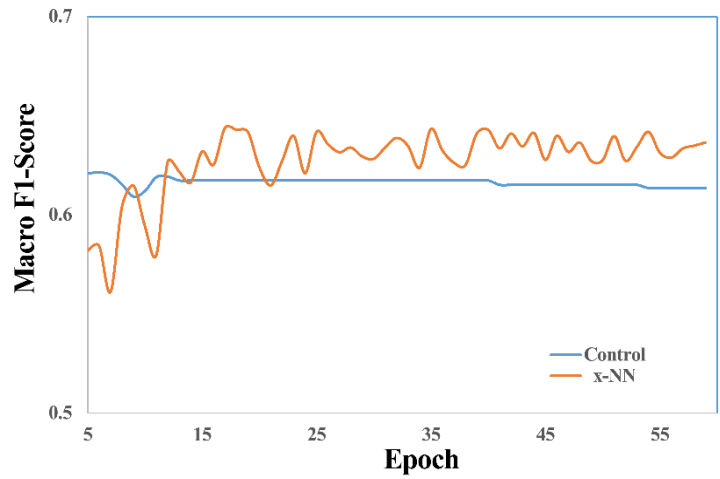

Figure 3. The Macro F1-Score of the *x*-NN network with Epoch on the dataset from Alzheimer's Disease Classification Technique Challenge PRCV 2021. It obviously that from epoch of ~15, the *x*-NN has a significantly higher accuracy than the control without fitting.

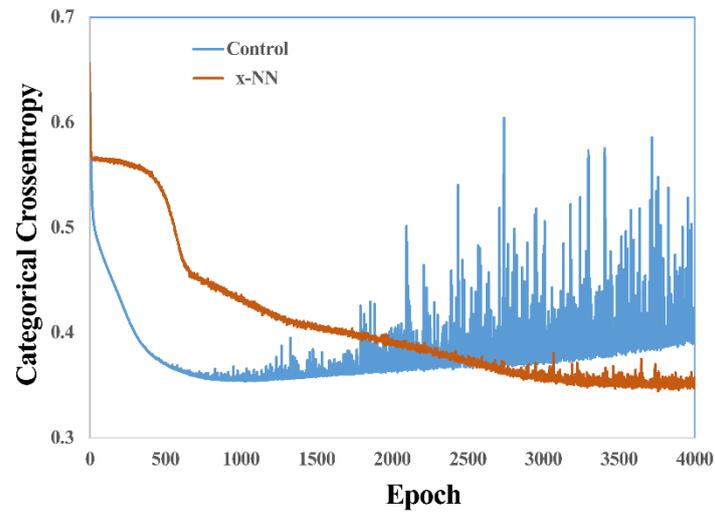

Figure 4. The categorical crossentropy loss of the *x*-NN network with Epoch on the dataset from COVID-19 - Clinical Data. It clearly that from epoch of ~2500, *x*-NN has a significantly lower loss than the control without fitting. Meanwhile, after the Epoch of 2000, the control fluctuates obviously but the LSnet is much more stable.

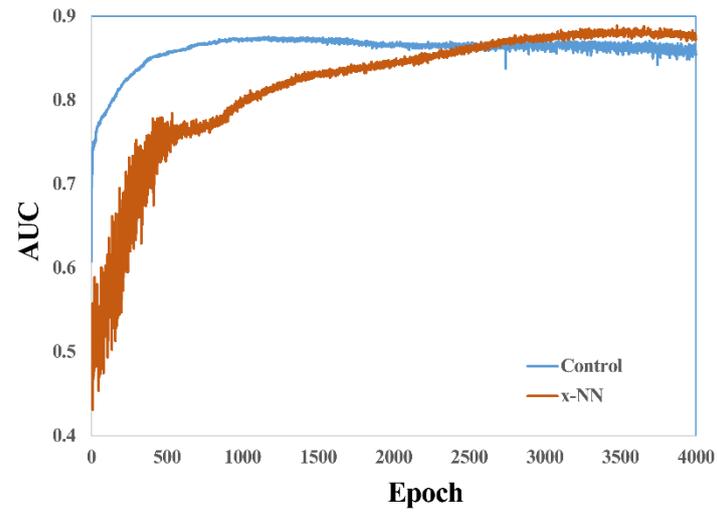

Figure 5. The AUC of the *x*-NN network with Epoch on the dataset from COVID-19 - Clinical Data. It clearly that from epoch of ~7500, *x*-NN has a significantly greater AUC without fitting.

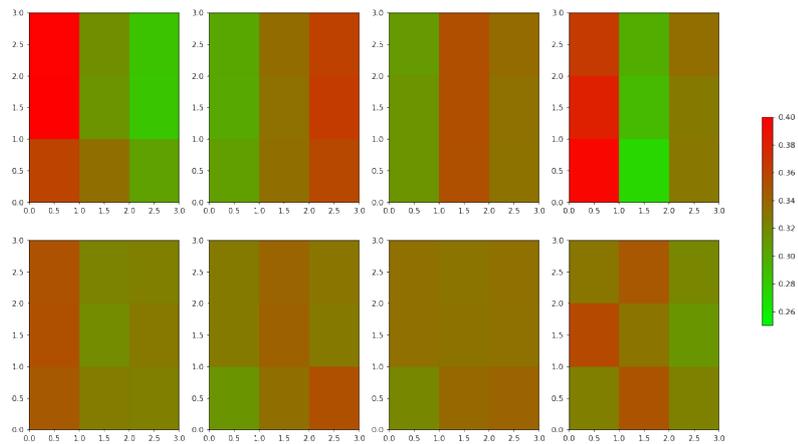

Figure 6. The attention distribution of the eight-head of the Multi-Head Attention Block in the in the task of the Alzheimer's Disease Classification Technique Challenge PRCV 2021. Each heatmap consists of a 3*3 heatmap block, where the *x*-axis and the *y*-axis represent the donors and the receptors of the attention, respectively. It is clearly that the heatmaps with the same *y*-axis are closer with each other and the attention is obviously related with the depth of the network, suggesting that the *x*-NN model can indeed adaptively allocate attention according the layers.